\documentclass[floatfix,amsmath,amsfonts,twocolumn]{revtex4}
\usepackage{enumitem}
\usepackage{amsmath}
\usepackage{graphicx}
\usepackage{epsfig}
\usepackage{bm}
\usepackage{bbold}
\usepackage{amssymb}
\usepackage{color}

\newcommand{\pt}{{\partial_t}}

\newcommand{\RE}{{\operatorname{Re}}}

\begin{document}

\title{Kerr nonlinearity effect on the stability of Wannier-Stark states in active optical systems}

\author{A. Verbitskiy}
\affiliation{School of Physics and Engineering, ITMO University, Kronverksky Pr. 49, bldg. A, St. Petersburg, 197101, Russia}

\author{A. Yulin}
\affiliation{School of Physics and Engineering, ITMO University, Kronverksky Pr. 49, bldg. A, St. Petersburg, 197101, Russia}

\date{\today}

\begin{abstract}
The paper provides an analytical and numerical investigation of the dynamics of a one-dimensional chain of coupled optical resonators with conservative cubic nonlinearity and the gain saturated by nonlinear losses. The linear dependency of the resonator eigenfrequencies on their indexes makes it possible to use Wannier-Stark states as lasing modes. Numerical simulations have shown that the dependency of the resonant frequencies on the light intensity strongly affects the stability of Wannier-Stark states. To explain the observed destabilization of monochromatic lasing based on Wannier-Stark states a simple perturbation theory has been developed and compared with the data obtained in the numerical simulations. 
\end{abstract}

\maketitle

\section{Introduction}

The concept of Wannier-Stark (WS) states originally developed in solid-state physics has been generalized for various physical systems. Initially, these states refer to electronic eigenstates localized within a periodic system under electric field \cite{discovery_1, discovery_2, discovery_3}. The interference of WS states may result in the appearance of Bloch oscillations (BOs), which represent a periodic movement of an electron \cite{discovery_4, discovery_5, discovery_6}. The similarity of the equations in quantum mechanics and optics predicted the formation of WS states for the light propagating in waveguide systems \cite{prediction_1, prediction_2}. Therefore, the optical analogues of BOs have been amply investigated \cite{BO_optics1, BO_optics2, BO_optics3, BO_optics4, BO_optics5, BO_optics6, BO_optics7, BO_optics8, BO_optics9}.

Experimental observation of optical BOs is less challenging compared to that in solid-state systems. Not surprising, after a theoretical study of optical WS states and BOs, they were both shortly discovered experimentally. The optical WS ladders were demonstrated in a modified Moire grating \cite{WS_exp1} and in superlattices of porous silicon \cite{WS_exp2}. In addition, WS states were obtained in photonic crystal lattices \cite{WS_exp3} and in curved waveguides \cite{WS_exp4}. BOs were also observed in waveguide arrays \cite{ BO_optics_exp1, BO_optics_exp2}, structures composed of porous silicon \cite{BO_optics_exp3, BO_optics_exp4}, circular systems \cite{BO_optics_exp5, BO_optics_exp6, BO_optics_exp7}, parity–time lattices \cite{BO_optics_exp8, BO_optics_exp9}, lasers \cite{laser}, plasmonic \cite{plasmonic1, plasmonic2, plasmonic3, plasmonic4, plasmonic5, plasmonic6} or exciton-polariton systems \cite{exciton1, exciton2, exciton3}. A detailed review on the phenomena described and the related effects was presented in \cite{BO_optics_review}. BOs appear to be quite common and thus occur in multiple areas of physics including atomic systems \cite{atomic1, atomic2, atomic3, atomic4, atomic5, atomic6}, coupled LC circuits \cite{LC}, and mechanical systems \cite{mech1, mech2, mech3, mech4}. Besides, WS states have been recently observed in such promising optical materials as perovskites \cite{interest_1}. The possibility of using WS localization in quantum technologies has been shown in \cite{interest_3}.

BOs represent a linear effect, and the important question is how the nonlinearities, always present in physical systems, affect the dynamics of WS states and, in particular, BOs. This challenge has been addressed in a number of works reporting the nonlinear shift and the disappearance of the resonant tunneling peaks \cite{nonl_WS1, nonl_WS2}, the shift of exceptional points \cite{nonl_WS3}, and the destruction of the WS beam profile in a photonic lattice \cite{nonl_WS4}. A set of papers have been focused on the nonlinear destabilization of BOs with a long-term field evolution \cite{nonl_BO1, nonl_BO2, nonl_BO3, nonl_BO4, nonl_BO5, nonl_BO6}. The understanding of the role the modulation instability plays in the destruction of BOs \cite{nonl_BO_expl} allowed to stabilize them by nonlinear control techniques \cite{nonl_BO_stab1, nonl_BO_stab2, nonl_BO_stab3, nonl_BO_stab4}. 
It is interesting to note that BOs in higher dimensional systems are often more robust than in one-dimensional ones \cite{nonl_BO_stab5, nonl_BO_stab6}, and could be used for resonant generation of new frequencies to name one possible application \cite{nonl_BO_rad_1}. BOs in nonlinear driven-dissipative systems have also been shown to demonstrate chaotic behavior \cite{nonl_BO_rad_2}. 

The existence of BOs and WS states in active optical systems, predicted in \cite{exciton1, exciton2, WS_state_lasing}, appears to be of fundamental and practical importance. The advantage of such systems is that BOs can remain there until the pump is switched off. In addition, fine tuning of the lasing frequency may be achieved by an appropriate heating of the sample, creating a temperature gradient in the chain of resonators \cite{BO_optics_exp1}. In terms of the prospects of WS states used as laser modes, their volume may be fairly large (the WS supermodes occupy many coupled resonators), which can increase the total lasing power. WS states can be regarded as internal localized modes within a superdimensional optical system. The resonant frequency gradient affects the width of the WS states and the density of their spectrum, so the inter-mode spectral distance can be made large enough to ensure that the gain pumps only one of the modes. The working frequency can be tuned by a geometrical shift of the pump spot, and so the tunability range is limited by the size of the whole system (provided that the gain is sufficiently broad band). The spectrum of the WS states is equidistant and all modes have the same spatial structure. These features can also be advantageous from the point of view of laser generation.

The well-known issue in the implementation of such systems is that they are based on optical microresonators which usually have a low quality factor. This means that lasing requires high gain, which is difficult to achieve experimentally. However, laser generation has recently been obtained in various microresonators \cite{microlasers_review_1, microlasers_review_2, microlasers_review_3, 
microlasers_review_5, microlasers_review_6}. The development of the concept of optical bound states in the continuum has already allowed to greatly improve the Q-factor of microresonators \cite{BIC_1, BIC_2}. On the other hand, the investigation of promising active media of perovskites resulted in a significant increase of the gain level achievable in solid-state lasers \cite{microlasers_review_4}. All the above factors can anticipate manufacturing of active optical systems supporting WS states in the near future. The possibility of experimentally observing persistent optical WS states increases the importance of theoretical investigation of such systems.    

The monochromatic lasing based on WS states was studied in our recent paper \cite{WS_state_lasing}, where it was shown that this regime of CW lasing can be achieved using the appropriate spatial profiling of the gain. However, the research neglected the effect of the nonlinear shift of resonator eigenfrequencies (Kerr nonlinearity). On the other hand, instantaneous Kerr nonlinearity is known to dramatically affect the stability of states (such nonlinearities of large magnitude occur, for example, in polariton systems \cite{polariton_laser}). The purpose of the current paper is to address the Kerr nonlinearity influence on the WS state lasing in a system of interacting optical cavities.

\begin{figure*}[ht]
 \begin{center} \includegraphics[width=0.8\textwidth]{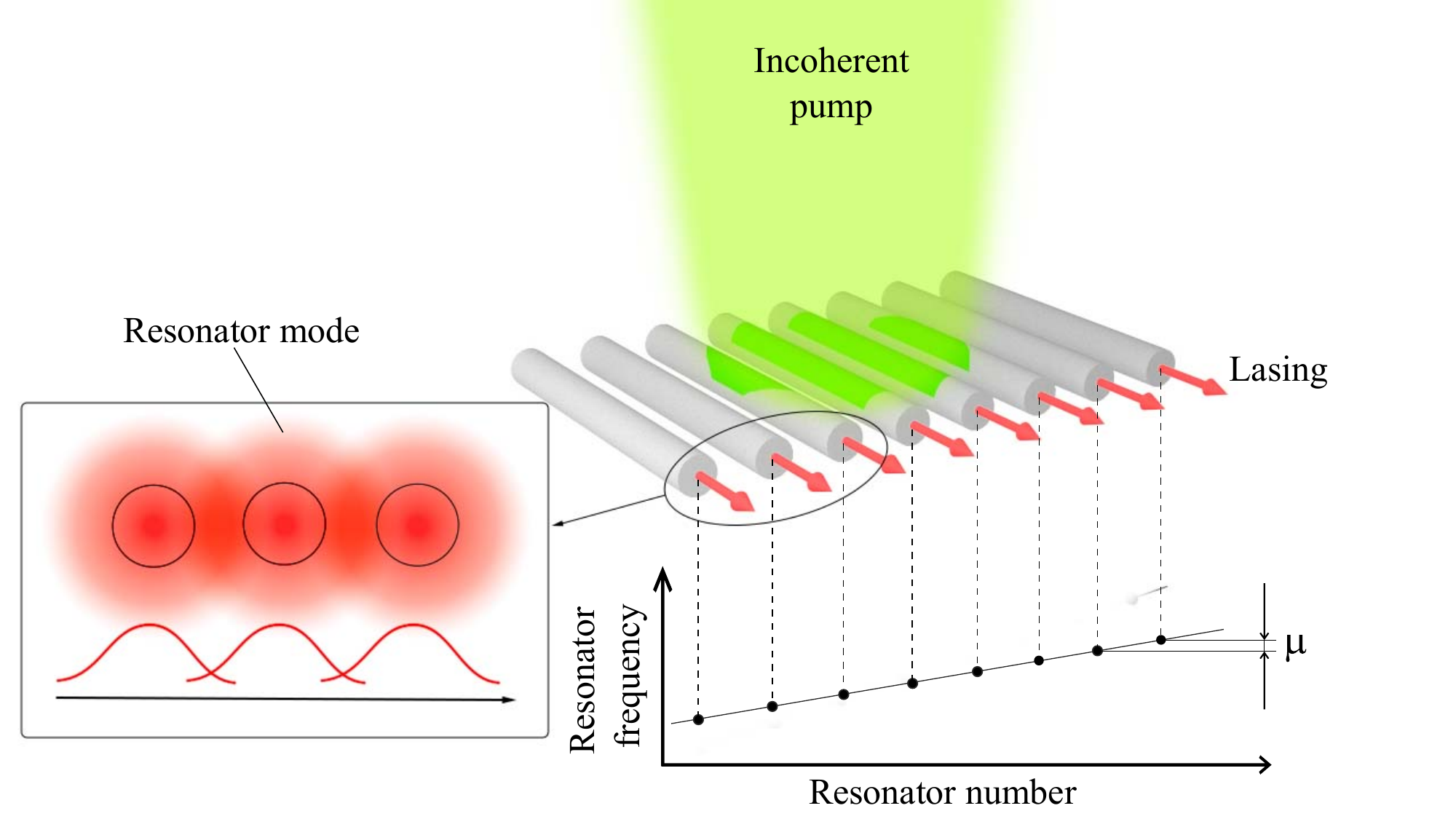}
  \caption{ One-dimensional array of coupled optical resonators (cylinders are shown in grey). The inset on the left is the zoomed area of resonators array. Red gradient circles show individual resonator eigenmodes (red curves below demonstrate their profiles). The overlap of the modes represents inter-resonator coupling allowing the waves to propagate in the system. The graph in the right bottom part illustrates the dependency of the eigenfrequencies of resonator modes on their indexes. $\mu$ is the frequency step between the fundamental modes of neighboring resonators. The optical pump (the green cone) is shed to the chain of the microresonators producing the lasing one (or more) modes of the system. The laser radiation leaving the resonators is shown by the red arrows.
  }
  \label{fig:one}
  \end{center}
\end{figure*}

The system is shown schematically in Fig.~\ref{fig:one}. The eigenfrequencies of the  resonator fundamental modes depend linearly on the resonator indexes. The coupling between the resonators appears because their modes are not perfectly confined inside them,
and thus the neighbouring resonators "talk to each other". Due to this feature the field in the whole system can be represented as a combination of WS states, which we will refer to as supermodes to distinguish them from the individual resonator modes. The resonators can be selectively pumped either optically or electrically so that positive gain is created in a group of resonators. 

The article is organized as follows. In Section II we introduce a mathematical model describing the system. Section III discusses the results of numerical studies of the stationary states forming in the system. Stability of the WS states is analytically considered in Section IV. To explain the observed numerical results, a simple perturbation theory is developed in Section V. Finally, in the Conclusion we briefly summarize the main findings of the work.

\section{Mathematical model of an array of optical resonators}

To describe an array of interacting nonlinear cavities we adopted a widely used model based on a generalized discrete nonlinear Schrodinger equation describing both the temporal evolution of the field in the coupled optical resonators and the propagation of monochromatic light in waveguide arrays \cite{Example_model1,Example_model2,Example_model3,Example_model4,Example_model5,Example_model6,Example_model7,Example_model8,Example_model9,Example_model10,Example_model11,Example_model12}. We assume that higher order modes can be neglected and account only for the fundamental ones. This approach can be justified in the case of the microresonators where the frequencies of the higher order modes are well detuned, so that only the fundamental modes see the positive gain created by incoherent pump. We also suppose that the nonlinear effects cannot result in any excitation of the higher order modes either. The coupling between the resonators is considered to be weak, and we can only account for next-neighbour coupling. The gain and the nonlinear effects are assumed to be small enough, and therefore, they can cause changes in field evolution with inverse characteristic time much less than the resonant frequency of the cavities. Then the dynamics of optical radiation in each of the resonators can be described by the equations for slow varying complex amplitudes $U_n(t)$ of the individual cavity modes 
\begin{widetext}
\begin{eqnarray}
i \pt U_n + \sigma (U_{n+1}+U_{n-1}) + \mu n U_n + \alpha|U_n|^2 U_n  + i \gamma_n U_n + i\beta_n|U_n|^2 U_n = 0,
	\label{eq:main}
\end{eqnarray}
\end{widetext}
where $n$ is the resonator number, $\sigma$ characterizes the interaction between the resonators, $\mu$ is the steepness of the linear dependence of the resonator eigenfrequencies on their index (the frequency difference between the neighbouring resonators), $\alpha$ is the coefficient of the Kerr nonlinearity, $\gamma_n$ are linear losses, and $\beta_n$ are nonlinear losses. 

Optical or electric pump can change the linear losses and even make them negative meaning that the individual resonator has positive gain and thus can switch into the lasing regime with the stationary amplitude determined by the balance of the linear and nonlinear losses. For the sake of completeness we consider the case when the nonlinear losses are a function of $n$. This can be achieved, for example, by additional absorbers build in some of the resonators. 
 
Throughout the paper we use the dimensionless parameters $\sigma=1$, $\mu=0.2$, $\gamma=0.01$, $\alpha=1$ and $\beta=1$. We see our paper as a proof of concept, so we have chosen the parameters convenient for theoretical demonstration of the effect. However, the parameters selected for the modelling do correspond to a real physical system, for example, to a chain of resonators having the quality factor $Q=10^3$ at the working wavelength $\lambda=825$~nm, the effective refracting coefficient $n_2=3 \cdot 10^{-18}$~m${^2}$~W$^{-1}$, the linear losses $\gamma = 3.6 \cdot 10^{11}$~s$^{-1}$ and the nonlinear coefficient $\alpha= 10^{10}$~J$^{-1}$, see \cite{BIC_lasing}. The coupling strength between the resonators and the frequency step between the fundamental modes of neighboring resonators were adopted as $\sigma =3.6 \cdot 10^{13}$~s$^{-1}$ and $\mu=0.72 \cdot 10^{13}$~s$^{-1}$. Nonlinear losses were selected as $\beta=10^{10}$~J$^{-1}$. 

In the conservative limit the eigenmodes of equation (\ref{eq:main}) are WS states $W_m(n)$, where $m$ is the index of the supermode, see \cite{BO_optics1}. The WS states have the properties of $W_{m+m^{\prime} }(n) = W_{m}(n+m^{\prime})$, and the eigenfrequency of the $m$-th WS state is given by $\omega_m=\mu m$. The WS states for our equation can be found analytically as $W_{m}(n)=J_{m-n}(\frac{2\sigma}{\mu})$.

The simplest way to excite a Bloch-type system based on an array of coupled optical resonators is single-site pumping, and it is only this latter case that we consider in the present research. In what follows, we consider the situation where the gain is created in the resonator with $n=0$. As has already been shown in \cite{WS_state_lasing}, in this case two WS supermodes with $m=\pm \tilde m$ ($\tilde m$ depends on $\mu$ and $\sigma$) 
see the same gain and thus grow at the same rate. To excite only one WS state in a controllable way it is sufficient to introduce some additional losses to a resonator where the intensity of one of two competing WS states is much larger than the intensity of the other one. This approach increases the increment of the second supermode, which suppresses the first one. In the current work we take the profile of the linear losses to be
\begin{eqnarray}
\gamma_n=\gamma, n \neq 0;-2 \tilde m, \notag
\\
\gamma_0=\gamma-a, \label{eq:lin_loss}
\\
\gamma_{-2 \tilde m}=4\gamma, \notag
\end{eqnarray}
where $\gamma$ stands for the linear losses in all resonators except the $n=0$ and $n=-2 \tilde m$. The losses in resonator $n=0$ are equal to $\gamma - a$, where $a$ is the linear gain, and so the losses can become negative. The parameter $a$ thus controls the effective losses seen by the mode. In contrast, the losses in resonator $n=-2 \tilde m$ are increased by factor of $4$.

\section{Formation of single and multi-frequency states}

In this section we discuss the numerical results of the stationary states' formation in the presence of the Kerr nonlinearity. All simulations shown in this paper have been done for the case when the gain and the nonlinear losses are present in one resonator only. We have also performed simulations for the situation where the nonlinear losses are present in all resonators, however, we did not observe any qualitative difference. In addition, the sign of the Kerr nonlinearity does not qualitatively affect the dynamics. Therefore, in this paper we discuss only the case of a positive $\alpha$ and the nonlinear losses present in the excited resonator only.  

\begin{figure*}[ht]
 \begin{center}  \includegraphics[width=0.9\textwidth]{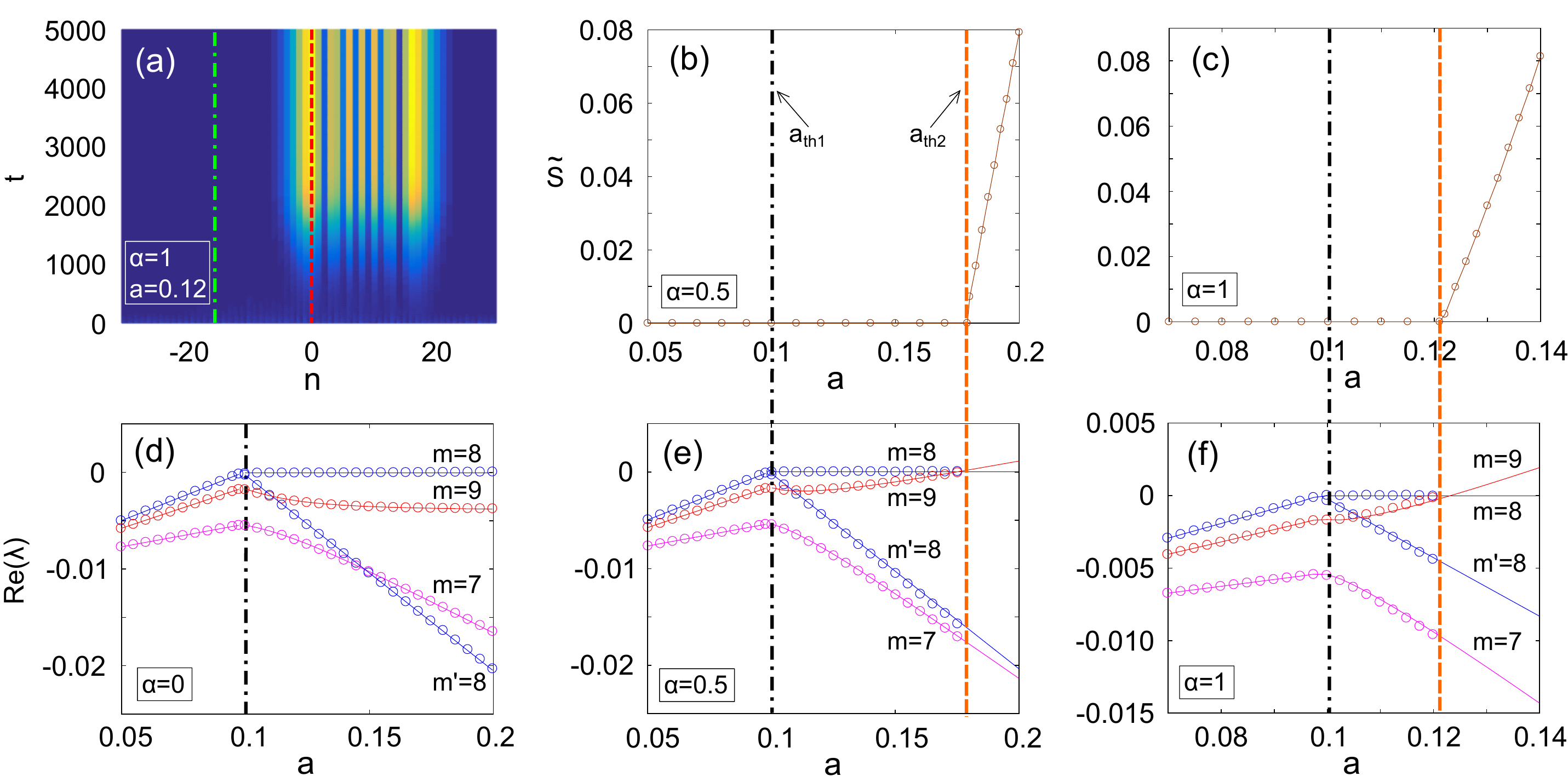}
  \caption{(Color online) (a) Evolution of the field amplitude $|U_n(t)|$ growing from weak noise for the Kerr nonlinearity coefficient $\alpha=1$ and the pump amplitude $a=0.12$. The red dashed line is the resonator with the gain, the green dash-dotted line shows the additional losses introduced for the mode selection. Panels (b) and (c) show the dependencies of the total field energy of the non-principal modes defined as $\tilde S=\int_{-\infty}^{\infty} S d\omega - \int_{\omega_{max}-\mu/2}^{\omega_{max} +\mu/2}S d\omega$ on the pump amplitude for $\alpha=0.5$ and $\alpha=1$ correspondingly. The frequency $\omega_{max}$ is the peak frequency of the most intense spectral line (the principal lasing mode). The bottom row shows the growth rates of Wannier-Stark (WS) states as functions of the pump amplitude for $\alpha=0$ (d), $\alpha=0.5$ (e) and $\alpha=1$ (f). The open circles are the real parts of the numerically found eigenvalues governing the dynamics of linear excitation against the background of the trivial state (before the lasing threshold marked by the black dash-dotted line) and against a single-frequency WS state (after the lasing threshold). To calculate the spectrum of the linear excitations we extract the spatial distribution of the stationary WS state from the direct numerical modelling. The solid lines show the corresponding dependencies found by the perturbation theory. The eigenvalues for eigenmodes are shown in blue (modes  $m=8$ and $m \prime=8$), red (modes $m=9$ and $m \prime=9$) and magenta (modes $m=7$ and $m \prime=7$). We use $m$ and $m'$ in the notation to distinguish the linear excitations of the progenitrix linear WS.
  }
  \label{fig:two}
  \end{center}
\end{figure*}

Direct numerical simulation of equation (\ref{eq:main}) showed that mode $m=8$ has the lowest threshold, and when parameter $a$ exceeds the threshold value $a_{th \, 1} \approx 0.1$ a nontrivial stationary state appears in the system. This state has a structure very much similar to that of WS state with $m=8$, because in this case both the dissipative terms and the conservative nonlinearity can be regarded as small perturbations not affecting the structure of the mode. The system thus switches to a monochromatic lasing state (see Fig.~\ref{fig:two}(a) illustrating the formation of this state).  

To check that the lasing is single-frequency we introduce the spectrum of the stationary state defined as $S(\omega)=\sum_n |\int_t^{t+T} U_n(t) e^{-i\omega t} dt|^2$, where $T$ is the time much larger than for any characteristic scales of the field dynamics. For a stationary state and a long integration time ($t$ and $T$ are large) the spectrum $S$ depends neither on $T$ nor on $t$. Then, we can calculate the $\tilde S=\int_{-\infty}^{\infty} S d\omega - \int_{\omega_{max}-\mu/2}^{\omega_{max} +\mu/2}S d\omega$, where $\omega_{max}$ is the frequency where the spectral intensity is maximum. It is $\mu$ that defines the spectral distance between the neighboring eigenstates for the linear conservative problem, and the dissipative and nonlinear effects do not significantly change the resonant frequencies of WS supermodes. Then we can refer to the lasing as single-frequency, if $\tilde S \approx 0$. The deviation from $0$ indicates that more than one WS states are excited in the system, and the spectrum contains more than one spectral line.

\begin{figure*}[ht]
 \begin{center}  \includegraphics[width=0.9\textwidth]{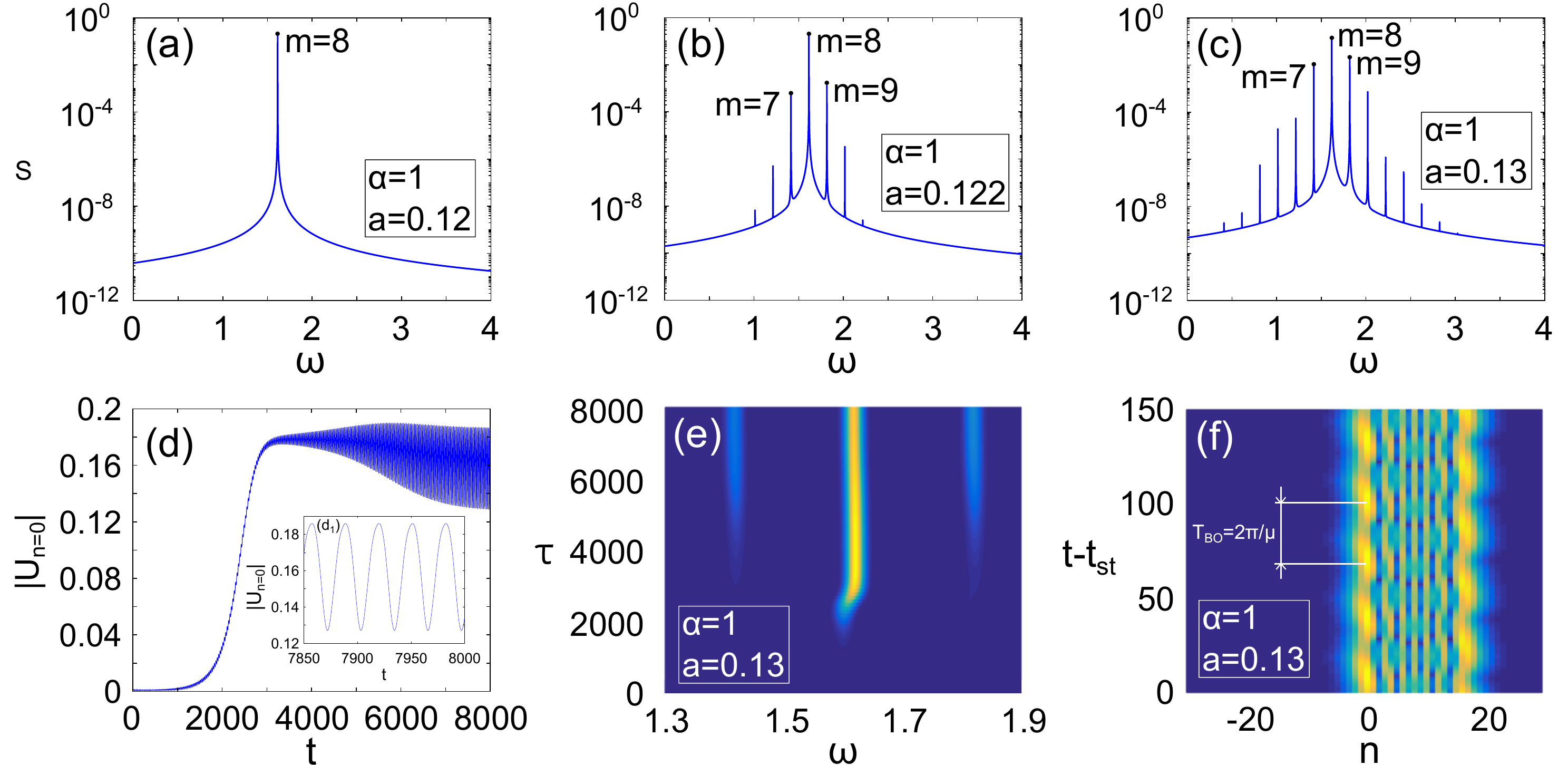}
  \caption{(Color online) Panels (a)-(c) show the temporal spectra of stationary states for different values of the pump: $a=0.12$, $a=0.122$ and $a=0.13$. $m$ is the index of Wannier-Stark supermode. Additional spectral lines appear after the second threshold $a \approx 0.122$ marked by the orange dashed line in Fig.~\ref{fig:two}(c, f). The increase of the pump enriches the spectrum making the sideband more intense. The background is non-zero   because the spectrum is calculated numerically in a large but finite temporal window. Panel (d) shows the numerically found evolution of the field absolute value $|U_{n=0}(t)|$ for the resonator $n = 0$ and the pump amplitude $a=0.13$. Inset ($d_1$) shows a zoomed part of the graph illustrating the stationary regime. To show the spectral evolution of the lasing modes we define the time-dependent spectrum as $S_W(\tau, \omega)=\sum_n | \int_{-\infty}^{\infty} U_n(t) g(t-\tau) \exp( -i\omega t) dt |^2$. The width $t_0$ of the function $g(t)=\exp(-t^2/{t_0}^2)$ refers to the width of the window where the spectrum is calculated, and the time $\tau$ refers to the position of the window. Panel (e) shows the temporal evolution of this spectrum for $a=0.13$, the window width is $t_0=150$. Panel (f) shows the spatial-temporal distribution of the field of the stationary multi-frequency state for $a=0.13$. In (f) $|U_n(t-t_{st})|$. Time $t_{st}$ is sufficient for the state to become stationary (in our simulations it is $8000$). For all simulations presented in Fig.~\ref{fig:three} the Kerr nonlinearity parameter is $\alpha=1$. The other parameters are the same as in Fig.~\ref{fig:two}.
  }
  \label{fig:three}
  \end{center}
\end{figure*}

The dependencies of $\tilde S$ on the parameter $a$ controlling the effective losses are shown in Fig.~\ref{fig:two}(b, c) for different values of the Kerr nonlinearity. At low pump levels $\tilde S$ is zero, but at the threshold value of $a$ the parameter $\tilde S$ starts growing. Thus, we can conclude that there is a second threshold pump $a_{th \, 2}$, at which single-frequency regime gets destroyed, and additional lines appear in the spectrum, see Fig.~\ref{fig:three}(a)-(c) showing the spectra below the second threshold (a), just above the second threshold (b) and at a relatively strong pump (c). Indeed, a single-frequency regime (a) changes to multi-frequency (b), and then the number of intense spectral lines increases with the pump amplitude (compare (b) and (c)). In the vicinity of the threshold $a_{th \, 2}$ the intensity of the sideband is orders of magnitude lower compared to the main harmonic, so from the field evolution in the coordinate domain it is difficult to distinguish single and multi-frequency state there. 

The value of the first threshold $a_{th \, 1}$ (the lasing threshold) does not depend on the strength of the Kerr nonlinearity $\alpha$. However, the second threshold, at which the lasing becomes multi-frequency does depend on $\alpha$, compare panels (b) and (c) of Fig.~\ref{fig:two}. When there is no Kerr nonlinearity we do not see a multi-frequency regime for the chosen parameters. With the increase of $\alpha$ the threshold $a_{th \, 2}$ decreases. In numerical simulations we observe that for large values of $\alpha$  the range of single-frequency regime becomes very small, but remains finite. Therefore, we can conclude that Kerr nonlinearity is crucial for the destruction of monochromatic solution.

To demonstrate how the multi-frequency regime appears we set our pump above the second threshold and assume a weak random noise as the initial condition. The temporal dynamics of the absolute value of the field is shown in  Fig.~\ref{fig:three}(d) for the resonator $n=0$. The field amplitude first grows exponentially without any noticeable oscillations. This behavior can be explained by the fact that the gain for only one WS state (with $m=8$) is above the lasing threshold, and so this mode grows until it is saturated by the nonlinear losses at a certain level (approximately at $t\approx 3000$). Then oscillations start developing until their amplitude stabilizes at approximately $t \approx 8000$. The oscillations of the amplitude absolute value can be explained by the beating of two (or more) modes having different frequencies, and thus, the appearance of the oscillations is a sign of switching to a multi-frequency regime.

The temporal spectrum of the field is calculated in a finite time window as follows 
\begin{eqnarray}
S_W(\tau, \omega)=\sum_n | \int_{-\infty}^{\infty} U_n(t) g(t-\tau) \exp( -i\omega t) dt |^2,
\label{spectr_window}
\end{eqnarray}
where $\tau$ is the position of the window and the function $g(t)$ defines the shape of the latter. We choose $g(t)=\exp(-t^2/{t_0}^2)$, where $t_0$ is the width of the window. Provided that the period of the oscillations $T_{oscil}$ and the amplitude variation time $T_{ampl}$ are very different, we can choose $T_{oscil} \ll t_0 \ll T_{ampl}$, and then the spectrum $S_W$ can be interpreted as that of oscillations at time $\tau$.

The evolution of the spectrum $S_W(\tau, \omega)$ is shown in Fig.~\ref{fig:three}(e). Initially, only one frequency grows, that corresponding to WS state with $m=8$. The width of the spectral lines is determined by that of the window, $t_0$. Then, at approximately $t=3000$ the Kerr nonlinearity comes to play, and thus, the spectral line shifts to higher frequencies (for the opposite sign of $\alpha$ the spectral line shifts to lower frequencies). 

At longer times ($t \approx 4500$) additional spectral lines become visible in the spectrum. At this point we see the developed oscillations of the absolute amplitude values in Fig.~\ref{fig:three}(d). Finally, a stationary multi-frequency state is formed, see panel (f) of Fig.~\ref{fig:three}. The intensity is periodically varying in time (see panel (f) and inset in panel (d) of Fig.~\ref{fig:three}) with the period equal to $\frac{2\pi}{\mu}$, and is similar to BOs.

The results described above allow to assume that the presence of the Kerr nonlinearity breaks the stability of the state containing only one WS supermode. In the next section, we discuss the numerical results on the stability of single-frequency WS states.

\section{Stability of Wannier-Stark states}

To check the hypothesis that a multi-frequency state appears as a result of the dynamical instability of a single-frequency state, we pose a spectral problem governing the stability of a stationary state. To do this, we find a solution for equation (\ref{eq:main}) in the form $U_n=(\tilde U_n +\psi_n) \exp(i\omega_{S}t)$, where $\psi_n$ is a small deviation of the field from its stationary solution $\tilde U_n$, $\omega_{S}$ is the frequency of the single-frequency stationary state. Substituting this ansatz into (\ref{eq:main}) and keeping only the terms linear with respect to $\psi_n$, we obtain the equation for $\psi_n$:
\begin{widetext}
\begin{eqnarray}
\pt \psi_n + i \omega_S \psi_n - i \sigma (\psi_{n+1}+\psi_{n-1}) - i \mu n \psi_n + \gamma_n \psi_n - 2 i \alpha |\tilde U_n|^2 \psi_n - i \alpha {\tilde U_n}^2 {\psi_n}^* + 2 \beta_n |\tilde U_n|^2 \psi_n + \beta_n {\tilde U_n}^2 {\psi_n}^* = 0.
	\label{eq:dyn_ampl_small_cor}
\end{eqnarray}
\end{widetext}

The solution for this equation can be found in the form $\psi_n = \phi_n \exp(\lambda t) + {\chi_n}^* \exp(\lambda^* t)$. The equations for $\phi_n$ and $\chi_n$ are
\begin{widetext}
\begin{eqnarray}
\lambda \phi_n = (-i \omega_S + i \mu n - \gamma_n + 2 i \alpha |\tilde U_n|^2 - 2 \beta_n |\tilde U_n|^2) \phi_n + i \sigma \phi_{n+1} + i \sigma \phi_{n-1} + ( i \alpha {\tilde U_n}^2 - \beta_n {\tilde U_n}^2) \chi_n,
	\label{eq:eigval_1}
 \\
\lambda \chi_n =  (-i \alpha [{\tilde U_n}^*]^2 - \beta_n [{\tilde U_n}^*]^2) \phi_n + (i \omega_S - i \mu n - \gamma_n - 2 i \alpha |\tilde U_n|^2 - 2 \beta_n |\tilde U_n|^2) \chi_n - i \sigma \chi_{n+1} - i \sigma \chi_{n-1}.
    \label{eq:eigval_2}
\end{eqnarray}
\end{widetext} 
The eigenvalue $\lambda$ defines the stability of the state: if $\RE (\lambda) > 0$, then the excitation $\psi_n$ grows in time, and thus, the stationary state is unstable. The eigenvectors $ (\phi_n, \chi_n)^T$ describe the structure of the eigenmodes existing against the background of the stationary solution. 

To analyze the spectral stability we need to know the single-frequency stationary solution. The latter as well as the frequency $\omega_S$ can be extracted from the results of numerical simulations for $a<a_{th \, 2}$. This approach allows us to find the eigenvalues governing the stability of the stationary state. The results are summarized in Fig.~\ref{fig:two}(d)-(f) showing the maximum real parts of $\lambda$'s (open circles) for various values of $\alpha$. 

Below the lasing threshold $a<a_{th \, 1}$ only a trivial state exists, and the spectrum of the linear excitations appears to be very similar to that of WS states (although not exactly the same because of the dissipative terms in the equation). Considering that the dissipative terms are small, we can relate each of the modes to the corresponding conservative WS state. Thus, in Fig.~\ref{fig:two}(d)-(f) the $\lambda$'s are marked by $m$'s, showing the index of the corresponding conservative WS mode.

For $a<a_{th \, 1}$, as shown in Fig.~\ref{fig:two}, the real part of $\lambda$ goes up with the increase of the pump. The spatial distribution of the gain and the losses results in the modes having the eigenvalues with different real parts, see \cite{WS_state_lasing} for details.   

The eigenvalue of the mode with $m=8$ has the largest real part and is the first to reach zero at $a=a_{th \, 1}$, making the trivial state unstable. The stationary state bifurcating from the trivial solution at $a_{th \, 1}$ is similar to the WS supermode having the largest increment (mode $m=8$ for our parameters).  

Since the background solution is nontrivial  for $a>a_{th \, 1}$ two modes with pure real eigenvalues exist. To mark the eigenvalues of the small excitations at $a>a_{th\, 1}$ we use the following notation. We can regard the eigenvalues as continuous functions of the pump $a$. The equations (\ref{eq:eigval_1})-(\ref{eq:eigval_2}) are factorized before the lasing threshold, and each of the solutions $\lambda$, $\phi=1, \chi=0$ and $\lambda^{*}$, $\phi=0, \chi=1$ correspond to the same physical state. In the conservative limit for the excitations against the trivial state the modes are characterized by the index $m$ (the index of the WS mode). We also use this index to mark the dissipative WS states, however, as we now have two eigenmodes we will denote the latter as $m$ and $m'$. When we study the case with the pump exceeding the lasing threshold, the two modes originating from the same linear WS mode will have different pure real values and different spatial structures. Thus, distinguishing between $m$ and $m'$ modes also has physical meaning.

Fig.~\ref{fig:two}(d)-(f) shows that the eigenvalue corresponding to the excitation with the structure similar to linear mode (with index $m=8$ in our case) splits into the zero eigenvalue, corresponding to the symmetry $U_n \rightarrow U_n e^{i \theta}$, and the second eigenvalue, that can be understood as the relaxation rate of the WS state amplitude to its equilibrium. The real part of the latter eigenvalue rapidly goes down with the increase of the pump. The other eigenvalues remain complex conjugated. However, it is necessary to acknowledge the hybridization of the modes. For example, if we represent the eigenvector of the mode $m=9$ as a series of different WS states, we can notice that the field has the dominating component with $m=9$, and the second largest component with $m=7$. The portion of the WS state with $m=7$ in the supermode structure increases with the pump $a$, resulting in stronger hybridization. 

Fig.~\ref{fig:two}(d) also shows that the real parts of all eigenvalues decrease (except the one equal to zero due to symmetry reasons). This behavior can be explained by the fact that the increase of the stationary state amplitude effectively increases the effective losses seen by the other modes. The exception is the mode $m=9$ (which is hybridized with the mode $m=7$): the real part of its eigenvalue decays slowly and remains nearly constant at relatively high pumps.  

The important fact is that in the presence of the Kerr nonlinearity the real part of the eigenvalue of the hybridized mode with $m=9$ starts growing at higher pumps and at $a=a_{th \, 2}$ becomes equal to zero, where our numerical simulations show the birth of the multi-frequency regime. The threshold $a_{th \, 2}$ is reached at smaller pump for the higher Kerr nonlinearities, for instance, for $\alpha=0.5$ the threshold is achieved at $a\approx 0.18$, and for $\alpha=1$ at $a\approx 0.12$.

Thus, we can suggest that the multi-frequency regime appears because of some hybridized mode growing, and that Kerr nonlinearity is crucial for this effect. To explain this situation we developed a simple perturbation theory presented in Section V.

\section{Perturbation theory}

To verify the hypothesis that the hybridization of the modes in the presence of the Kerr nonlinearity results in the destabilization of the single-frequency regime we developed a simple perturbation theory, assuming that the dynamics can be described in terms of time-dependent amplitudes of WS states $C_m$. In the simplest case we can neglect the effect of the dissipative and nonlinear terms on the eigenmodes' structure and find the solution in the form 
\begin{eqnarray}
U_n(t)=\sum_m C_m(t)W_{m}(n).
\label{eq:approx}
\end{eqnarray}
In the conservative limit the operator describing the interaction of the resonators is self-adjoined, and therefore, for the conservative WS states we have
$$ \sum_n W_{m_1}(n) W_{m_2}(n) = \delta_{m_1, m_2},$$
where $\delta_{m_1, m_2}$ is Kronecker symbol.

Substituting the expression (\ref{eq:approx}) into Eq. (\ref{eq:main}), multiplying the latter by $W_l$ and calculating the sum over $n$, we obtain the equations for $C_l$
\begin{widetext}
\begin{eqnarray}
\pt C_l =  i \mu l C_l - \sum_m  C_m \sum_n \gamma_n W_{m}(n) W_{l}(n) - \sum_{m_1} C_{m_1} \sum_{m_2} C_{m_2} \sum_{m_3} {C_{m_3}}^* \sum_n \beta_n W_{m_1}(n) W_{m_2}(n) W_{m_3}(n) W_{l}(n) + \notag
\\
+ i \alpha \sum_{m_1} C_{m_1} \sum_{m_2} C_{m_2} \sum_{m_3} {C_{m_3}}^* \sum_n W_{m_1}(n) W_{m_2}(n) W_{m_3}(n) W_{l}(n). 
	\label{eq:diss_nonl}
\end{eqnarray}
\end{widetext}

The next step is to find the stationary state. In the leading approximation order we can find the solution in the form $C_{l}=0$ for all $l$, except $l=\tilde m$. In other words, we find the solution in the form of just one excited WS state, which is justified, provided that the pump and the losses are weak enough. The equation for $C_{\tilde m}$ is then very simple 
\begin{widetext}
\begin{eqnarray}
\pt C_{\tilde m} =  i \mu {\tilde m} C_{\tilde m} - C_{\tilde m} \sum_n \gamma_n W_{\tilde m}^2(n) - C_{\tilde m} |C_{\tilde m}|^2 \sum_n \beta_n W_{\tilde m}^4(n) + i \alpha C_{\tilde m} |C_{\tilde m}|^2 \sum_n W_{\tilde m}^4(n). 
\label{eq:C8}
\end{eqnarray}
\end{widetext}

The stationary solution of (\ref{eq:C8}) can be found if the pump $a$ provides negative effective losses
\begin{eqnarray}
\sum_n \gamma_n W_{\tilde m}^2(n) <0.
\label{eq:cond_pert}
\end{eqnarray}
The amplitude of the stationary solution is 
\begin{eqnarray}
\tilde C_{\tilde m}(t) = \rho \exp (i \tilde \omega_{\tilde m}t),
\label{eq:st_ampl}
\end{eqnarray}
where $|\rho|=\sqrt{-\frac{\sum_n \gamma_n W_{\tilde m}^2(n)}{\sum_n \beta_n W_{\tilde m}^4(n)}}$, $\tilde \omega_{\tilde m}=\mu {\tilde m}+ \delta_{NL}$ and $\delta_{NL}  = \alpha  |\tilde C_{\tilde m}|^2 \sum_n W_{\tilde m}^4(n)$ is the nonlinear shift of the stationary state frequency. The amplitude $\rho$ can be set to be pure real without any loss of generality. The corrections to the single-mode solution can be found perturbatively, if needed.

Due to the presence of the term with $\alpha$, an increase of the WS state amplitude results in a frequency shift. Therefore, the spectral line in Fig.~\ref{fig:three}(e) shifts to the right, when the state intensity becomes high enough, and the excited state frequency is slightly different than the eigenfrequency of the linear mode.

We work in a regime where, due to mode competition \cite{WS_state_lasing}, it is only the mode having the maximum increment that grows. This mode modifies the pump in the way that the gain is saturated for all other modes. However, these modes can still start growing because of instantaneous Kerr nonlinearity resulting in parametric effects. Below we analyse the stability of the stationary solution.

We investigate the stability of a stationary solution of (\ref{eq:diss_nonl})
by writing an equation for small perturbation $d_l$ against a background of this solution. To find the solution of $C_l=\tilde C_l + d_l \exp(i \tilde \omega_{l} t)$ we get the equations
\begin{widetext}
\begin{eqnarray}
\pt d_l = \left( i (\mu l-\tilde \omega_{l}) -\Gamma_l \right) d_l  - \sum_{m} \eta_{l, m} d_{m} - \sum_{m} \nu_{l, m} d_{m}^{*},
\label{eq:dl_1}
\end{eqnarray}
where
$$\eta_{l, m}= 2 \sum_{m_1} \tilde C_{m_1} \sum_{m_2} {\tilde C_{m_2}^*} \sum_n (\beta_n-i\alpha) W_{m}(n) W_{m_1}(n) W_{m_2}(n) W_{l}(n),$$
$$\nu_{l, m}=\sum_{m_1} \tilde C_{m_1} \sum_{m_2} \tilde C_{m_2} \sum_n (\beta_n - i\alpha) W_{m_1}(n) W_{m_2}(n) W_{m}(n) W_{l}(n),$$
\end{widetext}
and $ \Gamma_l=\sum_n \gamma_n W_{l}^2(n)$.

This equations can be significantly simplified if we take the stationary solution in the leading approximation order, so in what follows $\tilde C_l=0$ for all $l \neq \tilde m$, and the amplitude $\tilde C_{\tilde m}$ is given by (\ref{eq:st_ampl}).  We also observed that WS states have different eigenfrequencies, and this frequency difference is supposed to be much greater than the coupling strength, which is true for sufficiently low state amplitude. This means that the coupling $\eta$ cannot result in efficient hybridization of the states, and we can neglect it. 

The term proportional to $\nu$ is different: it couples efficiently the states with the detunings of approximately the same absolute value but of the opposite signs (detuning is the frequency difference between the eigenfrequency of the state and the frequency of the stationary state). In other words, the coupling is efficient between the state $l_1$ and $l_2$ so that $\mu (l_1-l_2) \approx 2\tilde \omega_{\tilde m}$. For relatively weak amplitude of the state $\tilde \omega_{\tilde m} \approx \mu \tilde m$, this means that the coupling is efficient between the states $l_1=\tilde m - k$ and $l_{2}=\tilde m + k$. 

Then, accounting for the efficient coupling only, we obtain the equations for $d$ in the form
\begin{widetext}
\begin{eqnarray}
\pt d_{\tilde m+k} = \left( i (\mu (\tilde m +k)-\tilde \omega_{\tilde m}) -\Gamma_{\tilde m+k} - \eta_{{\tilde m +k}, {\tilde m +k}} \right) d_{\tilde m+k} -\nu_{{\tilde m+k}, {\tilde m-k}} d_{{\tilde m -k}}^{*}.
\label{eq:dl_2}
\end{eqnarray}
\end{widetext}
Physically, this means that in the leading approximation order, due to the four-wave mixing, the stationary WS state with $\tilde m$ couples two linear WS excitations with the indexes $l=\tilde m \pm k$.   

The eigenvalues governing the dynamics of each pair of the hybridized modes can easily be found
\begin{widetext}
\begin{eqnarray}
\lambda_{\tilde m \, k \, \pm} = i\mu k -\frac{ \Gamma_{\tilde m+k} + \Gamma_{\tilde m-k} + \eta_{\tilde m +k} +\eta_{\tilde m -k}^{*}   }{2} \pm \nonumber \\
\pm \sqrt{\left( i\delta_{NL}- \frac{ \Gamma_{\tilde m-k} - \Gamma_{\tilde m+k} + \eta_{\tilde m -k}^{*} - \eta_{\tilde m +k}   }{2}  \right)^2+ |\nu_{{\tilde m+k} \, {\tilde m-k}}|^2}.
\label{eq:eig_val_excitations}
\end{eqnarray}
\end{widetext}
Here, we have taken into account that $\nu_{m_1, m_2}=\nu_{m_2, m_1}$. 

Formula (\ref{eq:eig_val_excitations}) is derived for $k \neq 0$, however, it can easily be proved to work for $k=0$ as well. Indeed, in this case we can neglect the interaction of the excitations $m=\tilde m$ with all other linear excitations. Then from (\ref{eq:dl_1}) we obtain that
\begin{eqnarray}
\pt d_{\tilde m} = \left( -i \delta_{NL} -\Gamma_{\tilde m} - \eta_{\tilde m , \tilde m} \right) d_{\tilde m} -  \nu_{\tilde m, \tilde m} d_{\tilde m}^{*}.
\label{eq:dl_zero}
\end{eqnarray}
To find the solution in the form $( d_{\tilde m},  d_{\tilde m}^{*})^T\exp(\lambda_{\tilde m} t)$ we derive the quadratic equation for $\lambda_{\tilde m}$ 
\begin{widetext}
\begin{eqnarray}
(-i\delta_{NL} - \Gamma_{\tilde m} -\eta_{\tilde m , \tilde m} - \lambda_{\tilde m} )(i\delta_{NL} - \Gamma_{\tilde m} -\eta_{\tilde m , \tilde m}^{*} - \lambda_{\tilde m} ) = |\nu_{\tilde m, \tilde m}|^2,
\label{eq:ch_eq_zero}
\end{eqnarray}
\end{widetext}
having the solution
\begin{widetext}
\begin{eqnarray}
\lambda_{\tilde m \pm} = -\Gamma_{\tilde m} - \frac{\eta_{\tilde m , \tilde m}+\eta_{\tilde m , \tilde m}^{*}}{2} \pm \sqrt{ \left(i\delta_{NL}+\frac{\eta_{\tilde m , \tilde m}-\eta_{\tilde m , \tilde m}^{*}}{2} \right)^2 + |\nu_{\tilde m, \tilde m}|^2 }.
\label{eq:eig_val_excitationsZERO}
\end{eqnarray}
\end{widetext}

To compare the results of the perturbation theory with the numerical simulations we calculated the dependencies of the eigenvalues $\lambda_{\pm}$ on the pump intensities for the same parameters as those used in the numerical simulations. The dependencies of the real parts of the eigenvalues calculated perturbatively are shown in  Fig.~\ref{fig:two} by the solid lines for $\alpha=0$ (d), $\alpha=0.5$ (e) and $\alpha=1$ (f). Formulas (\ref{eq:eig_val_excitationsZERO}) and (\ref{eq:eig_val_excitations}) are shown to fit the data of numerical simulations very well.

The perturbation theory confirms that the excitations of the same structure as the stationary state do not result in any instability. On the other hand, coupling between the states with $l=\tilde m + k$ and $l=\tilde m - k$ can cause instability, resulting in the formation of a multi-frequency regime. For our parameters the instability is first generated by the modes with $k = \pm 1$, and Fig.~\ref{fig:two}(d)-(f) shows only their eigenvalues. However, the other pairs $|k|>1$ can also cause instability.

\section{Conclusion}
 
Wannier-Stark (WS) modes lasing was explored in the dissipative array of optical cavities with Kerr nonlinearity. We numerically demonstrated that in the presence of conservative nonlinearity there are various stationary regimes in Bloch-type system consisting of one-dimensional chain of resonators. Namely, there can be a single-frequency regime corresponding to the excitation of a single WS state. There can also be a multi-frequency regime resulting in the dynamics similar to non-decaying Bloch oscillations.

Numerical simulations showed that Kerr nonlinearity affecting the eigenfrequencies of the resonators can play a crucial role in switching from single-frequency to multi-frequency regime. The solution of the spectral problem showed that the regime is switched due to the instability giving rise of a number of WS states. 

To understand the destabilization mechanism of single-frequency regime we developed an analytical approach based on a simple perturbation theory. This analysis revealed that instability occurs due to parametric excitation of the modes with the indexes $l=\tilde m \pm k$. It is the modes with $k=\pm 1$ that have the highest increment, and so oscillations appear on the frequencies detuned by parameter $\mu$ (frequency difference between the neighbouring resonators) from the frequency of the initially excited state. 

To summarize, we investigated numerically and analytically the dynamics and stability of optical WS states in a resonator system with dissipation and nonlinearity. The advantage of such a system is the tunability of the radiation frequency and the possibility to increase the radiation power by increasing the working mode volume. Our findings appear to be prospective for further fundamental research and can be potentially used to achieve stable monochromatic generation in microlaser arrays.

\begin{acknowledgments}
This work was supported by the Ministry of Science and Higher Education of the Russian Federation, Goszadanie no. 2019-1246.  
\end{acknowledgments}


\begin{thebibliography}{99} 

\bibitem{discovery_1} G. H. Wannier, Elements of Solid State Theory, Cambridge
University Press, London, 1959, pp. 190–193.

\bibitem{discovery_2} Shockley, William. "Stark ladders for finite, one-dimensional models of crystals." Physical Review Letters 28.6 (1972): 349.

\bibitem{discovery_3} Emin, David, and C. F. Hart. "Existence of Wannier-Stark localization." Physical Review B 36.14 (1987): 7353.

\bibitem{discovery_4} F. Bloch, Uber die Quantenmechanik der Elektronen in Kristallgittern, Z. Phys. 52, 555 (1929). 

\bibitem{discovery_5} C. Zener, A theory of the electrical breakdown of solid dielectrics, Proc. R. Soc. A 145, 523 (1934).

\bibitem{discovery_6}  W.V. Houston, Acceleration of electrons in a crystal lattice, Phys. Rev. 57, 184 (1940).


\bibitem{prediction_1} Monsivais, Guillermo, Marcelo del Castillo-Mussot, and Francisco Claro. "Stark-ladder resonances in the propagation of electromagnetic waves." Physical review letters 64.12 (1990): 1433.

\bibitem{prediction_2} De Sterke, C. Martijn, John E. Sipe, and Laura A. Weller-Brophy. "Electromagnetic Stark ladders in waveguide geometries." Optics letters 16.15 (1991): 1141-1143.


\bibitem{BO_optics1} U. Peschel, T. Pertsch,  and F. Lederer,  Optical Bloch oscillations in waveguide arrays. Opt. Lett. 23, 1701 (1998).

\bibitem{BO_optics2} A. Kavokin, G. Malpuech, A. Di Carlo, P. Lugli, and F. Rossi, Photonic Bloch oscillations in laterally confined Bragg mirrors, Phys. Rev. B 61, 4413 (2000). 

\bibitem{BO_optics3} G. Malpuech, A. Kavokin, G. Panzarini, and A. Di Carlo, Theory of photon Bloch oscillations in photonic crystals, Phys. Rev. B 63, 035108 (2001).

\bibitem{BO_optics4} S Longhi, Optical Zener-Bloch oscillations in binary waveguide arrays, Europhys. Lett. 76 416 (2006).

\bibitem{BO_optics5} Efremidis, Nikolaos K., and Demetrios N. Christodoulides. "Bloch oscillations in optical dissipative lattices." Optics letters 29.21 (2004): 2485-2487.

\bibitem{BO_optics6} Lenz, G., I. Talanina, and C. Martijn De Sterke. "Bloch oscillations in an array of curved optical waveguides." Physical Review Letters 83.5 (1999): 963.

\bibitem{BO_optics7} Peschel, Ulf, Christoph Bersch, and Georgy Onishchukov. "Discreteness in time." Open Physics 6.3 (2008): 619-627.

\bibitem{BO_optics8} Wilkinson, P. B. "Photonic Bloch oscillations and Wannier-Stark ladders in exponentially chirped Bragg gratings." Physical Review E 65.5 (2002): 056616.

\bibitem{BO_optics9} Longhi, Stefano. "Bloch oscillations in complex crystals with P T symmetry." Physical review letters 103.12 (2009): 123601.


\bibitem{WS_exp1} De Sterke, C. Martijn, et al. "Observation of an optical Wannier-Stark ladder." Physical Review E 57.2 (1998): 2365.

\bibitem{WS_exp2} Ghulinyan, Mher, et al. "Zener tunneling of light waves in an optical superlattice." Physical review letters 94.12 (2005): 127401.

\bibitem{WS_exp3} Qi, Xinyuan, et al. "Observation of accelerating Wannier–Stark beams in optically induced photonic lattices." Optics Letters 39.4 (2014): 1065-1068.

\bibitem{WS_exp4} Mukherjee, Sebabrata, et al. "Modulation-assisted tunneling in laser-fabricated photonic Wannier–Stark ladders." New journal of physics 17.11 (2015): 115002.


\bibitem{BO_optics_exp1} T. Pertsch, P. Dannberg, W. Elflein, A. Brauer, and
F. Lederer, Optical Bloch Oscillations in Temperature
Tuned Waveguide Arrays, Phys. Rev. Lett. 83, 4752
(1999).

\bibitem{BO_optics_exp2} Morandotti, R., et al. "Experimental observation of linear and nonlinear optical Bloch oscillations." Physical Review Letters 83.23 (1999): 4756.

\bibitem{BO_optics_exp3} Sapienza, Riccardo, et al. "Optical analogue of electronic bloch oscillations." Physical Review Letters 91.26 (2003): 263902.

\bibitem{BO_optics_exp4} V. Agarwal, J. A. del Rio, G. Malpuech, M. Zamfirescu,
A. Kavokin, D. Coquillat, D. Scalbert, M. Vladimirova,
and B. Gil, Photon Bloch Oscillations in Porous Silicon
Optical Superlattices, Phys. Rev. Lett. 92, 097401 (2004).

\bibitem{BO_optics_exp5} S. Longhi, M. Lobino, M. Marangoni, R. Ramponi, P. Laporta, E. Cianci, and V. Foglietti, Semiclassical motion
of a multiband Bloch particle in a time-dependent field:
Optical visualization, Phys. Rev. B 74 155116 (2006).

\bibitem{BO_optics_exp6} Usievich, B. A., et al. "A circular system of coupled waveguides and an optical analogue of Bloch oscillations." Optics and spectroscopy 97 (2004): 790-795.

\bibitem{BO_optics_exp7} Chiodo, Nicola, et al. "Imaging of Bloch oscillations in erbium-doped curved waveguide arrays." Optics letters 31.11 (2006): 1651-1653.

\bibitem{BO_optics_exp8} Regensburger, Alois, et al. "Parity–time synthetic photonic lattices." Nature 488.7410 (2012): 167-171.

\bibitem{BO_optics_exp9} Ye-Long Xu, W.S. Fegadolli, Lin Gan, Ming-Hui Lu,
Xiao-Ping Liu, Zhi-Yuan Li, A. Scherer and Yan-Feng
Chen, Experimental realization of Bloch oscillations in
a parity-time synthetic silicon photonic lattice, Nature
Communications volume 7, Article number: 11319 (2016). 


\bibitem{laser} S. Longhi, Dynamic localization and Bloch oscillations in the spectrum of a frequency mode-locked laser, Opt. Lett. 30, 786 (2005).



\bibitem{plasmonic1} A.R. Davoyan, I.V. Shadrivov, A.A. Sukhorukov, and Y.S. Kivshar, Plasmonic Bloch oscillations in chirped metal-dielectric structures, Appl. Phys. Lett. 94, 161105 (2009).

\bibitem{plasmonic2} V. Kuzmiak, S. Eyderman, and M. Vanwolleghem, Controlling surface plasmon polaritons by a static and/or time-dependent external magnetic field, Phys. Rev. B, 86, 045403 (2012).

\bibitem{plasmonic3} Bo Han Cheng, Yi-Chieh Lai, and Yung-Chiang Lan, Plasmonic Photonic Bloch Oscillations in Composite Metal–Insulator–Metal Waveguide Structure, Plasmonics, 9, 137 (2014).

\bibitem{plasmonic4} V. Kuzmiak, A. A. Maradudin, and E. R. Mendez, Surface plasmon polariton Wannier–Stark ladder, Opt. Lett. 39, 1613 (2014),

\bibitem{plasmonic5} A. Block, C. Etrich, T. Limboeck, F. Bleckmann, E. Soergel, C. Rockstuhl and S. Linden, Bloch oscillations in plasmonic waveguide arrays, Nature Communications, volume 5, Article number: 3843 (2014). 

\bibitem{plasmonic6} H. Wetter, Z. Fedorova, and S. Linden, Observation of the Wannier–Stark ladder in plasmonic waveguide arrays, Optics Letters, 47, 12, 3091 (2022).


\bibitem{exciton1} H. Flayac, D. D. Solnyshkov, and G. Malpuech, Bloch oscillations of an exciton-polariton Bose-Einstein condensate, Phys. Rev. B 83, 045412 (2011).

\bibitem{exciton2} H. Flayac, D. D. Solnyshkov, and G. Malpuech, Bloch oscillations of exciton-polaritons and photons for the generation of an alternating terahertz spin signal, Phys. Rev. B, 84, 125314 (2011).

\bibitem{exciton3} J. Beierlein, O.A. Egorov, T.H. Harder, P. Gagel, M. Emmerling, C. Schneider, S. Hofling, U. Peschel, and S. Klembt, Bloch Oscillations of Hybrid Light-Matter Particles in a Waveguide Array, Adv. Opt. Mater. 9, 2100126 (2021).

\bibitem{BO_optics_review} I.L. Garanovich, S. Longhi, A.A. Sukhorukov, and Y.S. Kivshar, Light propagation and localization in modulated photonic lattices and waveguides
Physics Reports 518 1 (2012).


\bibitem{atomic1} M.B. Dahan, E. Peik, J. Reichel, Y. Castin, and C. Salomon, Bloch Oscillations of Atoms in an Optical Potential, Phys. Rev. Lett. 76, 4508 (1996).

\bibitem{atomic2} S. Wilkinson, C. Bharucha, K. Madison, Q. Niu, and M. Raizen, Observation of Atomic Wannier-Stark Ladders in an Accelerating Optical Potential, Phys. Rev. Lett. 76, 4512 (1996).

\bibitem{atomic3} H. R. Zhang and C. P. Sun, Bloch oscillations of polaritons of an atomic ensemble in magnetic fields, Phys. Rev. A 81, 063427 (2010).

\bibitem{atomic4} Z.A. Geiger, K.M. Fujiwara, K. Singh, R. Senaratne, S.V. Rajagopal, M. Lipatov, T. Shimasaki, R. Driben, V.V. Konotop, T. Meier, and D.M. Weld, Observation and Uses of Position-Space Bloch Oscillations in an Ultracold Gas, Phys. Rev. Lett. 120, 213201 (2018).

\bibitem{atomic5} Z. Pagel, W. Zhong, R.H. Parker, C.T. Olund, N.Y. Yao, and H. Muller, Symmetric Bloch oscillations of matter waves, Phys. Rev. A 102, 053312 (2020).

\bibitem{atomic6} L. Masi, T. Petrucciani, G. Ferioli, G. Semeghini, G. Modugno, M. Inguscio, and M. Fattori, Spatial Bloch Oscillations of a Quantum Gas in a “Beat-Note” Superlattice, Phys. Rev. Lett. 127, 020601 (2021).


\bibitem{LC} S. Bahmani and A.N. Askarpour, Bloch oscillations and Wannier-Stark ladder in the coupled LC circuits, Phys. Lett. A 384, 126596 (2020).


\bibitem{mech1} G. Monsivais and R. Esquivel-Sirvent, Stark Ladder Resonances in Acoustic Waveguides, Journal of Mechanics of Materials and Structures, 2, 8, 1585 (2007).

\bibitem{mech2} G. Monsivais, R. Mendez-Sanchez, A. de Anda, J. Flores, L. Gutierrez, and A. Morales, Elastic Wannier–Stark Ladders in Torsional Waves
Journal of Mechanics of Materials and Structures, 2, 1629 (2007).

\bibitem{mech3} N. Lanzillotti-Kimura, A. Fainstein, B. Perrin, B. Jusserand, O. Mauguin, L. Largeau, and A. Lemaitre, Bloch Oscillations of THz Acoustic Phonons in Coupled Nanocavity Structures, Phys. Rev. Lett. 104, 197402 (2010).

\bibitem{mech4} Y.-K. Liu, H.-W. Wu, P. Hu, and Z.-Q. Sheng, Spatial Bloch oscillations in acoustic waveguide arrays, Appl. Phys. Express 14, 064501 (2021).


\bibitem{interest_1} Berghoff, Daniel, et al. "Low-field onset of Wannier-Stark localization in a polycrystalline hybrid organic inorganic perovskite." Nature communications 12.1 (2021): 1-9.

\bibitem{interest_3} Karamlou, Amir H., et al. "Quantum transport and localization in 1d and 2d tight-binding lattices." npj Quantum Information 8.1 (2022): 1-8.


\bibitem{nonl_WS1} Wimberger, Sandro, et al. "Nonlinearity-induced destruction of resonant tunneling in the Wannier-Stark problem." Physical Review A 72.6 (2005): 063610.

\bibitem{nonl_WS2} Wimberger, Sandro, Peter Schlagheck, and Riccardo Mannella. "Tunnelling rates for the nonlinear Wannier–Stark problem." Journal of Physics B: Atomic, Molecular and Optical Physics 39.3 (2006): 729.

\bibitem{nonl_WS3} Elsen, Christoffer, et al. "Exceptional points in bichromatic Wannier–Stark systems." Journal of Physics B: Atomic, Molecular and Optical Physics 44.22 (2011): 225301.

\bibitem{nonl_WS4} Qi, Xinyuan, et al. "Observation of accelerating Wannier–Stark beams in optically induced photonic lattices." Optics Letters 39.4 (2014): 1065-1068.

\bibitem{nonl_BO1} D. Cai, A.R. Bishop, and N. Gronbech-Jensen, Electric-Field-Induced Nonlinear Bloch Oscillations and Dynamical Localization, Phys. Rev. Lett 74, 1186 (1995).

\bibitem{nonl_BO2} R. Morandotti, U. Peschel, J.S. Aitchison, H.S. Eisenberg, and Y. Silberberg,  Experimental observation of linear and nonlinear optical Bloch oscillations, Phys. Rev. Lett., 83, 4756 (1999).

\bibitem{nonl_BO3} O. Morsch, J.H. Muller, M. Cristiani, D. Ciampini, and E. Arimondo, Bloch oscillations and mean-field effects of Bose-Einstein condensates in 1D optical lattices,  Phys. Rev. Lett. 87, 140402 (2001).

\bibitem{nonl_BO4} M. Cristiani, O. Morsch, J.H. Muller, D. Ciampini, and E. Arimondo, Experimental properties of Bose-Einstein condensates in one-dimensional optical lattices: Bloch oscillations, Landau-Zener tunneling, and mean-field effects. Phys. Rev. A 65, 063612 (2002).

\bibitem{nonl_BO5} M. Gustavsson, E. Haller, M.J. Mark, J.G. Danzl, G. Rojas-Kopeinig, and H.-C. Nagerl,  Control of Interaction-Induced Dephasing of Bloch Oscillations,  Phys. Rev. Lett. 100, 080404 (2008).

\bibitem{nonl_BO6} Y.V. Bludov, V.V., Konotop, and M. Salerno, Dynamical localization of gap-solitons by time periodic forces. EPL (Europhys. Lett.) 87, 20004 (2009).

\bibitem{nonl_BO_expl} V.V. Konotop and M. Salerno, Modulation instability in Bose-Einstein condensates in optical lattices, Phys. Rev. A 65, 021602 (2002).

\bibitem{nonl_BO_stab1} M. Salerno, V.V. Konotop and Y.V. Bludov, Long-living Bloch oscillations of matter waves in periodic potentials, Phys. Rev. Lett. 101,
30405 (2008).

\bibitem{nonl_BO_stab2} Y.V. Bludov, V.V. Konotop, and M. Salerno, Linear superpositions of nonlinear matter waves in optical lattices,  EPL (Europhys. Lett.) 93, 30003 (2011).

\bibitem{nonl_BO_stab3} C. Gaul, R.P.A. Lima, E. Diaz, C.A. Muller, and F. Domnguez-Adame,  Stable Bloch oscillations of cold atoms with time-dependent interaction, Phys Rev. Lett. 102, 255303 (2009).

\bibitem{nonl_BO_stab4} Y.V. Bludov, V.V. Konotop, and M. Salerno, Long-lived matter wave Bloch oscillations and dynamical localization by time-dependent nonlinearity management,  J. Phys. B 42, 105302 (2009).

\bibitem{nonl_BO_stab5} R. Driben, V.V. Konotop, T. Meier, and A.V. Yulin, Bloch oscillations sustained by nonlinearity, Sci. Rep. 7, 3194 (2017).

\bibitem{nonl_BO_stab6} Khomeriki, Ramaz. "Solitonic Bloch oscillations in two-dimensional optical lattices." Physical Review A 82.3 (2010): 033816.

\bibitem{nonl_BO_rad_1} A. Yulin, R. Driben and T. Meier,Bloch oscillations and resonant radiation of light propagating in arrays of nonlinear fibers with high-order dispersion,  Phys. Rev. A 96, 033827 (2017).

\bibitem{nonl_BO_rad_2} Verbitskiy, Alexey, Alexey Yulin, and A. G. Balanov. "Chaotic Bloch oscillations in dissipative optical systems driven by a periodic train of coherent pulses." Physical Review A 107.5 (2023): 053519.


\bibitem{WS_state_lasing} Verbitskiy, A., and A. Yulin. "Excitation of Wannier-Stark states in a chain of coupled optical resonators with linear gain and nonlinear losses." JETP, 138.4, (2024).

\bibitem{microlasers_review_1} Yang, Xi, et al. "Fiber optofluidic microlasers: structures, characteristics, and applications." Laser \& Photonics Reviews 16.1 (2022): 2100171.

\bibitem{microlasers_review_2} Chen, Zhi, et al. "Emerging and perspectives in microlasers based on rare-earth ions activated micro-/nanomaterials." Progress in Materials Science 121 (2021): 100814.

\bibitem{microlasers_review_3} Zhukov, A. E., et al. "Quantum-dot microlasers based on whispering gallery mode resonators." Light: Science \& Applications 10.1 (2021): 1-11.

\bibitem{microlasers_review_5} Sumetsky, M. "Optical bottle microresonators." Progress in Quantum Electronics 64 (2019): 1-30.

\bibitem{microlasers_review_6} He, Huajun, et al. "MOF‐Based Organic Microlasers." Advanced Optical Materials 7.17 (2019): 1900077.


\bibitem{BIC_1} Koshelev, Kirill, et al. "Subwavelength dielectric resonators for nonlinear nanophotonics." Science 367.6475 (2020): 288-292.

\bibitem{BIC_2} Hwang, Min-Soo, et al. "Nanophotonic nonlinear and laser devices exploiting bound states in the continuum." Communications Physics 5.1 (2022): 106.


\bibitem{microlasers_review_4} Zhang, Qing, et al. "Halide perovskite semiconductor lasers: materials, cavity design, and low threshold." Nano Letters 21.5 (2021): 1903-1914.


\bibitem{polariton_laser} Bajoni, Daniele, et al. "Polariton laser using single micropillar GaAs- GaAlAs semiconductor cavities." Physical review letters 100.4 (2008): 047401.



\bibitem{Example_model1} W. Deering, M. Molina, and G. Tsironis, Directional couplers with linear and nonlinear elements, Appl. Phys. Lett. 62, 2471, (1993).

\bibitem{Example_model2} J. Eilbeck, G. Tsironis, and S. K. Turitsyn, Stationary states in a doubly nonlinear trimer model of optical couplers, Phys. Scr. 52, 386 (1995).

\bibitem{Example_model3} P.G. Kevrekidis, K.O. Rasmussen and A.R. Bishop, The discrete nonlinear Schr\"{o}dinger Equation: a survey of recent results, International Journal of Modern Physics B 15, 2833 (2001).

\bibitem{Example_model4} N.K. Efremidis, S. Sears, D.N. Christodoulides, J.W. Fleischer, M. Segev, Discrete solitons in photorefractive optically induced photonic lattices, Phys. Rev. E 66,  046602, (2002).

\bibitem{Example_model5} A.A. Sukhorukov, Yu.S. Kivshar, H.S. Eisenberg, Y. Silberberg, Spatial optical solitons in waveguide arrays, IEEE J. Quantum Electron., 39,  31–50, (2003).

\bibitem{Example_model6} U. Peschel, O. Egorov, and F. Lederer, Discrete cavity solitons, Opt. Lett. 29, 1909, (2004). 

\bibitem{Example_model7} D. Pelinovsky and P. G. Kevrekidis, Stability of discrete solitons in nonlinear Schrödinger lattices, Physica D 212, 1-19, (2005).

\bibitem{Example_model8} A. Yulin, D. V. Skryabin, and P. St. J. Russell, Dissipative localized structures of light in photonic crystal films, Opt. Express, 13, 3529, (2005). 

\bibitem{Example_model9} A.V. Yulin, D.V. Skryabin , A.G. Vladimirov, Modulational instability of discrete solitons in coupled waveguides with group velocity dispersion, Opt. Express , 14, 12347, (2006).

\bibitem{Example_model10} H. Susanto, P. G. Kevrekidis, B. A. Malomed, R. Carretero-Gonzalez, and D. J. Frantzeskakis, Discrete surface solitons in two dimensions, Phys. Rev. E 75, 056605, (2007).

\bibitem{Example_model11} O.A. Egorov and F. Lederer, Lattice-cavity solitons in a degenerate optical parametric oscillator , Phys. Rev. A 76, 053816, (2007).

\bibitem{Example_model12} O.A. Egorov,  F. Lederer, and Yu.S. Kivshar, How does an inclined holding beam affect discrete modulational instability and solitons in nonlinear cavities?, Opt. Express 15, 4149 (2007).


\bibitem{BIC_lasing} Mylnikov, Vasilii, et al. "Lasing action in single subwavelength particles supporting supercavity modes." ACS nano 14.6 (2020): 7338-7346.

\end{thebibliography}
\end{document}